\documentstyle[11pt,psfig]{article}
\newcommand{\kms}{km\,s$^{-1}$}

\title{{\bf Conversion Problems:\\ How (Not) to Determine Molecular Masses in
    Dwarf Galaxies}}
\author{S.~H\"uttemeister$^1$\\
\vspace{0.1cm}\\
\normalsize $^1$Astronomisches Institut der Ruhr-Universit\"at, Postfach 102148, D-44780 Bochum, Germany\\
}
\date{}
\begin{document}
\maketitle
\def\bull{\vrule height .9ex width .8ex depth -.1ex}
\makeatletter
\def\ps@plain{\let\@mkboth\gobbletwo
\def\@oddhead{}\def\@oddfoot{\hfil\tiny
``Dwarf Galaxies and their Environment'';
Bad Honnef, Germany, 23-27 January 2001; Eds.{} K.S. de Boer, R.-J.Dettmar, U. Klein; Shaker Verlag}%
\def\@evenhead{}\let\@evenfoot\@oddfoot}
\makeatother


%
\section{Molecular Gas Masses and the Conversion Factor $X_{\rm CO}$}
The determination of molecular gas masses in star forming dwarf irregular 
galaxies is crucial to assess the star formation process in dwarfs: 
Does it proceeed in the same way as in giant spiral galaxies, or are e.g.\
star formation efficiencies different? How large is the gas reservoir in
dwarf galaxies, i.e.\ are dwarfs rated as `starbursts' really 
undergoing a burst of activity or are they able to continue forming
stars at the present rate for a significant fraction of the Hubble Time? 
Star formation in gas rich dwarfs also provides us with a
local analogy to conditions in the high-redshift universe, since
within the context of the hierarchical bottom-up scenario of galaxy formation
we expect the building blocks of larger galaxies to have been small,
metal-poor (low $Z$) systems similar to present-day dwarf galaxies. 

However, the derivation of the molecular gas content of dwarf galaxies has
been a long-standing problem. CO, as the only practical tracer of cold 
molecular gas, has been (and to some extent still is) notoriously diffucult 
to detect. Yet, star formation clearly takes place in many dwarf irregulars 
and Blue Compact Dwarf Galaxies, as is shown in detail by investigations of 
their stellar populations.

Taylor, Kobulnicky \& Skillman (1998) have argued that there is a threshold 
metallicity of $12 + \log({\rm O/H}) \sim 7.9$ below which CO is not 
detectable. This may be understood in terms of models of the structure of 
low metallicity molecular clouds suggested by e.g.\ Madden et al.\ (1997) 
and Bolatto, Jackson \& Ingalls (1999): CO is present only in the cloud core,
which is surrounded by an H$_2$ envelope where carbon is in the form of CI
or CII. In this `hidden H$_2$' scenario, the CO core may vanish entirely while
molecular gas still is abundant if $Z$ is very low. 

The (in)famous conversion factor $X_{\rm CO} = I({\rm CO})/N({\rm H}_2)$
allows an easy estimate of the molecular gas mass based on measuring just
the $^{12}$CO(1--0) transition. $X_{\rm CO}$ has been calibrated in the disk
of the Milky Way. Even there, values vary between 
$1.6 \cdot 10^{20}$\,cm$^{-2}$(K\,\kms )$^{-1}$ (Hunter et al.\ 1997) and 
$3 \cdot 10^{20}$\,cm$^{-2}$(K\,\kms )$^{-1}$ (commonly used in the 80's).
While it is likely that the `correctly' calibrated value -- if it exists -- 
is slightly below $2 \cdot 10^{20}$\,cm$^{-2}$(K\,\kms )$^{-1}$, a 
`Standard' Conversion Factor (SCF) of 
$2.3 \cdot 10^{20}$\,cm$^{-2}$(K\,\kms )$^{-1}$ (Strong et al.\ 1988) is
widely used. 

The existence of a SCF can be motivated somewhat heuristically assuming that
(a) CO counts clouds, (b) the emission is optically thick and thermalized
and (c) the individual clouds are virialized, which implies a size-linewidth
relation. Then, a relation of the form $X_{\rm CO} = {\rm const}\ 
n({\rm H}_2)^{1/2}\ T_{\rm kin}^{-1}$ is obtained. If the molecular ISM is
dominated by `typical' molecular clouds with similar (average) densities 
and temperatures, $X_{\rm CO}$ can be expected to be constant. In the 
`hidden H$_2$' scenario outlined above, it should, however, not be surprising 
if a different (larger) value of $X_{\rm CO}$ is found.

\section{The Virial Method -- and its Drawbacks}
To determine $X_{\rm CO}$, an independent way of measuring the cloud
mass (or, equivalently if the distance to the source is known, the H$_2$ 
column density) is needed. The most obvious such method is the calculation
of viral masses, by using $M_{\rm vir} [{\rm M_{\odot}}]  = 190\ \Delta v^2\ 
[{\rm km/s}]\ D/2 [{\rm pc}]$. The constant depends slightly on the assumed 
density profile of the cloud. In practice, both the cloud diameter $D$ and
the velocity width $\Delta v$ can be obtained from high resolution CO 
observations. 

It can, of course, be debated whether molecular clouds in dwarf galaxies 
are in fact virialized. A size-linewidth relation consistent with the one
found for the Milky Way and nearby spirals seems to hold for large ($M \geq 
10^5 {\rm M}_{\odot}$) Giant Molecular Clouds (GMCs) in nearby
dwarfs, e.g.\ NGC\,1569, IC\,10 or the SMC (Taylor et al.\ 1999). 
While virialization does not necessarily follow from this (a similar relation
is valid for cirrus clouds which are clearly not in virial equilibrium), the 
assumption is at least not obviously erroneous. The situation is more critical 
for more distant dwarf galaxies like NGC\,4214 (Walter et al.\ 2001) or 
Haro\,2 (Fritz 2000). In these cases, the resolution, even with 
interferometers, is too poor to resolve GMCs -- the structures seen are 
Giant Molecular Associations (GMAs) which do not follow the size-linewidth 
relation and are likely not virialized. If this complication is
ignored, $M_{\rm vir}$ is unlikely to be a true measure of 
$M_{\rm cloud}$. In addition, the linewidth can be affected by systematic 
streaming motions or rotation, external pressure, magnetic fields, a stellar 
or atomic gas component, etc. 

Virial masses have been used to determine $X_{\rm CO}$ for a number of dwarf
galaxies. The values found vary significantly, from $(6.6 \pm 1.5)$\,SCF
in NGC\,1569 (which may be slightly too high, Taylor et al.\ 1999) to
$(1.2 \pm 0.7)$\,SCF in NGC\,4214 (Walter et al. 2001). For Haro\,2, 
Fritz (2000) derives $(2.9 \pm 0.8)$\,SCF if the CO(1--0) line is used and 
$(1.2 \pm 0.6)$\,SCF for the higher resolution observations in the
(2--1) transition. `Typical' values from the literature for a number of 
dwarfs, including IC\,10, give $(2 ... 3)$\,SCF; these form the basis of
the general conclusion that $X_{\rm CO}$ in dwarfs is somewhat higher than the
SCF. All galaxies for which results are given here have comparable 
metallicities. 

Are the differences that are derived real, or due to (systematic) errors?
The `hidden H$_2$' should be largely included in the virial masses, at least
as long as size and linewidth of the observed CO complex are dominated by the 
relative movement of small CO cores within a larger bound complex. Given that
the typical sizes of the resolved units exceed 100\,pc, this is likely. 

\begin{figure}
\begin{center}
\psfig{file=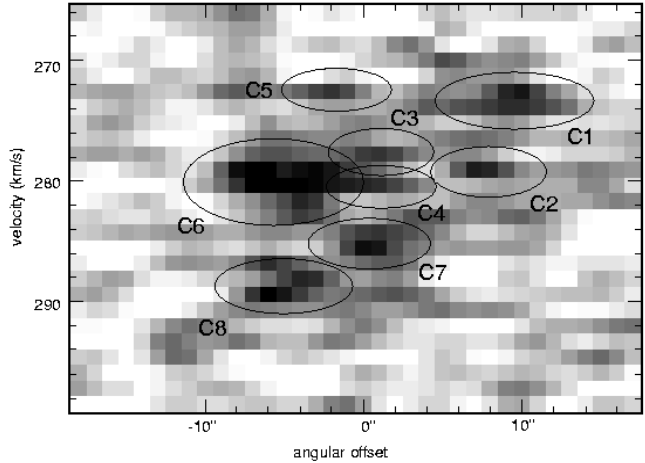,width=9cm}
\vspace*{-5cm}
\hspace*{8cm}
{\parbox{4cm}{
\caption{\small
An example of the `deconvolution' of a CO complex into clumps in
NGC\,4214, taken from Walter et al.\ (2001). It is obvious that the process
is highly dependent on resolution and the cutoff criteria employed.}}}
\end{center}
\end{figure}

The dependence of $M_{\rm vir}$ on the resolution is an obvious, but quite
possibly decisive, problem. Fig.\,1 illustrates the deconvolution of a
molecular complex in NGC\,4214 into subunits for which virial masses are
derived. It is evident that there is a strong dependence on the velocity 
resolution, which in this case is very high. Since the square of the velocity 
width enters $M_{\rm vir}$, a lower velocity resolution would increase 
the virial mass significantly and increase the low value of $X_{\rm CO}$
found for NGC\,4214. The two different values given for Haro\,2  
clearly illustrate the influence of linear resolution on $X_{\rm CO}$.

>From this, it may seem that the high values of $X_{\rm CO}$ usually 
derived for dwarf galaxies could be artifacts of too low a velocity or linear
resolution. However, we find a high $X_{\rm CO}$ for the nearby NGC\,1569,
where the resolution is good. Also, some unresolved clumps in NGC\,4214
yield $X_{\rm CO} < {\rm SCF}$, which is unexpected since in an unresolved 
clump $D$ should be an upper limit, leading to an overestimate of the virial
mass. Thus, resolution cannot explain all the variation in $X_{\rm CO}$
found in dwarfs of similar metallicity. It is, however, an issue that has to
be considered carefully. 

Even more basic are the varying definitions of a `cloud' that are employed.
Is it correct to always use the smallest unit that can be resolved? At some
point, we will either miss the `hidden H$_2$' envelopes and just regard the
single CO cores, and/or we will use subunits that are moving within a larger
bound entity. In the first case, we will clearly underestimate $X_{\rm CO}$,
in the second case, $D$ will be too small, while $\Delta v$ may be too large,
with an uncertain outcome for $M_{\rm vir}$. It is likely that with a 
resolution of $\sim 100$\,pc, this stage is not yet reached, but the very
small values of $X_{\rm CO}$ for unresolved clumps in NGC\,4214 serve as
a warning. Even the methods employed for size determinations vary from
using e.g.\ the FWHM, the 90\% of flux contour or sophisticated deconvolution
routines (e.g.\ based on information entropy, Fritz 2000). 

Given the uncertainties of the virial masses, it is interesting to examine 
another, seemingly entirely independent, method to derive $X_{\rm CO}$. 

\section{The Radiative Transfer Method -- and its Drawbacks}

If at least three CO transitions, including at least one of a rare isotopomer
like $^{13}$CO, are measured, a simple radiative transfer model can be 
employed to derive the average properties of the emitting gas. For a given 
density, metallicity, isotope ratio (and velocity gradient, in the commonly 
used LVG approximation of radiative transfer), $X_{\rm CO}$ is obtained as 
$X_{\rm CO} = n({\rm H}_2)/({\rm grad}v\ T_{\rm MB})$. Fig.\,2 illustrates 
the general principle.

\begin{figure}
\begin{center}
\psfig{file=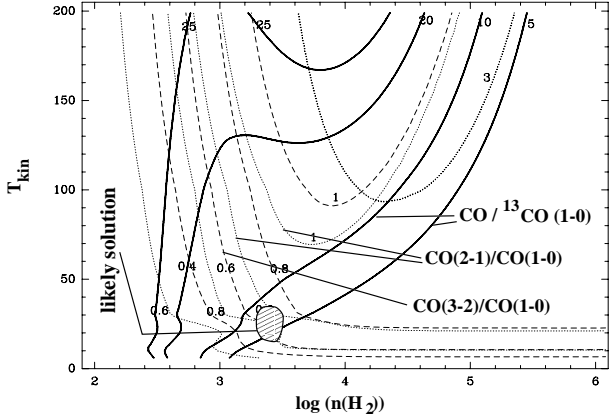,width=9cm}
\vspace*{-5.3cm}
\hspace*{8cm}
{\parbox{6cm}{
\caption{\small
The principle of determining a solution for the cloud properties
from CO line ratios in a non-LTE radiative transfer model. A solution is
found from the intersection of curves corresponding to observed line ratios.
Since ratio curves involving only $^{12}$CO transitions are almost parallel,
at least one $^{13}$CO line needs to be measured.}}}
\end{center}
\end{figure}

The method is less independent of the virial method than it might seem since
it implicitly assumes a size-linewidth relation (see Fritz 2000 for details).
A major drawback is observational, however: the $^{13}$CO transition(s) 
essential for the analysis are extremely faint and difficult to measure with
present instruments, especially since the quality of the spectra should be 
high enough to obtain reliable line ratios. Thus, such measurements exist
for only very few dwarfs, and they are confined to the 
molecular peak positions which may not be typical of the ISM properties in
general. 

The most crucial limitation of the method is the {\em extremely local} nature
of the value of $X_{\rm CO}$ it yields. Only the CO-emitting cores 
are taken into account, since the (average) local properties of the CO
emission itself are analysed. The envelopes (i.e.\ the hidden H$_2$) are
{\em not} included. Since the properties within the cores can be expected
to be similar to those of molecular clouds in metal rich galaxies, a value
for $X_{\rm CO}$ that is close to the SCF may be expected -- and is indeed
found for Haro\,2 ($X_{\rm CO} = (0.7 ... 1.3)$\,SCF, Fritz 2000). Note,
however, that this is not inconsistent with what is derived using the virial 
method at high resolution. 

Even simple one-component radiative transfer models show that $X_{\rm CO}$ can
vary by a factor of $>10$ {\em only depending on excitation}, without changing
the metallicity at all. If $Z$ is decreased, the local $X_{\rm CO}$
for a given $n/T$ increases slightly, but the changing size of the CO-emitting
core is likely to dominate this effect. In contrast, $X_{\rm CO}$ {\em
drops significantly} in a low density, high temperature environment, as is
qualitatively expected from the $\sim n^{1/2}/T_{\rm kin}$ relation. This
situation is not likely to be relevant for low-$Z$ dwarfs since this
`diffuse' gas component is likely to be devoid of CO or dissociated 
entirely. This is confirmed by the `disk-GMC-like' properties derived for the 
CO emission in Haro\,2, suggesting fairly low temperatures and high 
densities. Still, the detections of CO(3--2) emission in a number of dwarfs
(see the contribution by M\"uhle et al.\ in these preceedings) show that
extended, warm molecular gas is present (and detectable) in star forming 
dwarf galaxies. Exact densities and temperatures for this component remain
to be determined.  

The `diffuse gas' scenario is very important for metal-rich starburst systems,
where $X_{\rm CO}$ is found to be lower than the SCF by a factor of 3 ... 8.

\section{Final Considerations}
In summary, the general picture of the `correct' way of determining molecular
masses in dwarf galaxies remains somewhat clouded. Clearly, an analytical
{\em metallicity-X$_{CO}$-relation} of the form $\log X_{\rm CO}/{\rm SCF} =
a + b(12 + \log[{\rm O/H}])$ with $a = 5.9 ... 9.3$ and $b = -0.65 ... -1.0$
as proposed by e.g.\ Arimoto, Sofue \& Tsojimoto (1996) is close to useless
for any individual galaxy. It does describe the general trend correctly, but
with a scatter $> 5$ for a given $Z$, it misses all complications and 
individualities. 

Despite the strong possibility of systematic errors in the determination of
$X_{\rm CO}$ in dwarfs, some variation is very likely to be real. Even the
generally elevated value of $X_{\rm CO}$ in dwarfs compared to the SCF is not
undisputed, but we need to carefully eliminate possible sources of error 
before drawing far-reaching conclusions. Most prominently, we need to consider
how `global' the $X_{\rm CO}$ we determine is -- even with the virial method, 
only one molecular complex at a time
is probed, and the conversion factor derived may well differ in different
places in a given galaxy. Showing the dependence of $X_{\rm CO}$ on factors
other than $Z$ (e.g.\ the UV-field, cosmic rays, the gas morphology, pressure,
magnetic fields etc.) is indeed an important goal of further studies. 
Clearly, $X_{\rm CO}$ can never be a `constant' to be blindly applied 
to `determine' molecular gas masses. Instead, it needs to be probed by all
possible means and can then be a valuable {\em diagnostic} of ISM properties.

{\small
\begin{description}{} \itemsep=0pt \parsep=0pt \parskip=0pt \labelsep=0pt
\item {\bf References}

\item
Arimoto, N., Sofue, Y., Tsojimoto, T. 1996, PASJ 48, 275
\item
Bolatto, A.D., Jackson, J.M., \& Ingalls, J.D. 1999, ApJ 513, 275 
\item
Fritz, T. 2000, PhD Thesis, Bonn University
\item
Hunter, S.D., Bertsch, D.L., Catelli J.A., et al.\ 1997, ApJ 481, 205
\item
Madden, S.C., Poglitsch, A., Geis, N., Stacey, G.J., Townes, C.H. 1997,
ApJ 483, 200
\item
Strong, A.W., Bloemen, J.B.G.M., Dame, T.M.. et al.\ 1988, A\&A 207, 1
\item
Taylor, C.L., Kobulnicky, H.A., \& Skillman, E.D. 1998, AJ 116, 2746
\item
Taylor, C.L., H\"uttemeister, S., Klein, U., Greve, A. 1999, A\&A 349, 424
\item
Walter, F., Taylor, C.L., H\"uttemeister S., Scoville, N.Z., McIntyre, V.
2001, AJ, in press (astro-ph/0011098)

\end{description}
}

\end{document}